\documentclass[]{aastex631}
\usepackage{tabularx}
\usepackage{amsmath}
\usepackage{longtable}
\usepackage{multirow}
\usepackage{enumitem} \setlist[itemize]{noitemsep, topsep=4pt}

\begin{document}

\title{The Sonora Substellar Atmosphere Models. V: A Correction to the Disequilibrium
Abundance of CO$_2$ for Sonora Elf Owl}

\author[0000-0002-0413-3308]{Nicholas F. Wogan}
\affiliation{NASA Ames Research Center, Moffett Field, CA 94035}

\author[0000-0001-5864-9599]{James Mang}
\altaffiliation{NSF Graduate Research Fellow.}
\affiliation{Department of Astronomy, University of Texas at Austin, Austin, TX 78712, USA}

\author[0000-0003-1240-6844]{Natasha E. Batalha}
\affiliation{NASA Ames Research Center, Moffett Field, CA 94035}

\author{Kevin Zahnle}
\affiliation{NASA Ames Research Center, Moffett Field, CA 94035}

\author[0000-0003-1622-1302]{Sagnick Mukherjee}
\affiliation{Department of Astronomy and Astrophysics, University of California, Santa Cruz, CA 95064, USA}
\affiliation{Department of Physics and Astronomy, Johns Hopkins University, Baltimore, MD, USA}

\author[0000-0001-6627-6067]{Channon Visscher}
\affiliation{Center for Exoplanetary Systems, Space Science Institute, Boulder, CO 80301, USA}
\affiliation{Chemistry \& Planetary Sciences, Dordt University, Sioux Center, IA 51250, USA}

\author[0000-0002-9843-4354]{Jonathan J. Fortney}
\affiliation{Department of Astronomy and Astrophysics, University of California, Santa Cruz, CA 95064, USA}

\author[0000-0002-5251-2943]{Mark S. Marley}
\affiliation{Lunar \& Planetary Laboratory, Department of Planetary Sciences, University of Arizona, Tucson, AZ 85721, USA}

\author[0000-0002-4404-0456]{Caroline V. Morley}
\affiliation{Department of Astronomy, University of Texas at Austin, Austin, TX 78712, USA}

\begin{abstract}
  To aid the interpretation of observations of substellar atmospheres, \citet{Mukherjee2024} created the Sonora Elf Owl grid of model atmospheres, simulations that accounted for disequilibrium quench chemistry. However, Sonora Elf Owl did not accurately estimate CO$_2$ quenching because the models quenched the gas with respect to the full atmosphere equilibrium, but CO$_2$ should have instead been quenched with respect to the disequilibrium (i.e., quenched) abundance of CO. As a result, Sonora Elf Owl under-predicted the CO$_2$ abundance by several order of magnitude in some instances, an amount that JWST is sensitive to. Here, we release version two of the Sonora Elf Owl grid which has corrected CO$_2$ concentrations. Additionally, in version two we remove PH$_3$ as a spectral contributor since our spectra consistently contained too much PH$_3$ absorption. The new spectra can be found as an update to the original Zenodo postings.
\end{abstract}

\section*{}

The Elf Owl atmosphere model grid set
\citep{Mukherjee2024} of the Sonora substellar model family \citep{Marley2021} covers a wide range of effective temperatures, gravities, metallicities, carbon-to-oxygen ratios and vertical mixing coefficients ($K_{zz}$). The grid captures chemical disequilibrium effects, namely, the quenching or ``freeze-out'' of chemical reactions as gases mix from the hotter deep atmosphere to the cooler upper atmosphere. Quenching determines the mid- to upper-atmosphere concentration of major atmospheric species, like CH$_4$, CO$_2$, CO, NH$_3$, that are detectable spectroscopically with the James Webb Space Telescope (JWST) and other observatories.

\citet{Mukherjee2024} accounted for disequilibrium chemistry by using quench parametrizations developed by \citet{Zahnle2014}. \citet{Zahnle2014} used a kinetics and diffusion code that tracked hundreds of elementary chemical reactions to determine the timescale of conversion between various atmospheric species. For example, they found that the characteristic timescale for CO to CH$_4$ conversion is approximated by
\begin{equation} \label{eq:coch4}
  \tau_\mathrm{CO} = 3 \times 10^{-6} P^{-1} \exp(42000/T)
\end{equation}
where $\tau_\mathrm{CO}$ is in seconds, $P$ is bars, and $T$ is in Kelvin. This conversation timescale can be compared to a timescale of vertical atmospheric mixing:
\begin{equation}
  \tau_\mathrm{mix} = H^2/K_{zz}
\end{equation}
Here, $\tau_\mathrm{mix}$ is the timescale of vertical mixing in seconds, $H$ is the pressure scale height in cm, and $K_{zz}$ is the eddy diffusion coefficient in cm$^2$/s. In the deep, hot atmosphere reactions are fast compared to transport ($\tau_\mathrm{CO} \ll \tau_\mathrm{mix}$) and equilibrium can be maintained between relevant gases as they are vertically mixed. As gases are transported upward to lower pressures and temperature, reaction timescales lengthen and ultimately become similar-to or larger than the vertical mixing timescale (e.g., $\tau_\mathrm{CO} \ge \tau_\mathrm{mix}$). At this point, kinetics fail to maintain equilibrium between relevant species and gas concentrations become ``frozen'' or quenched.

\citet{Mukherjee2024} used these principles, and the \citet{Zahnle2014} chemical timescales, to try to account for CO$_2$, CH$_4$, CO, NH$_3$, N$_2$, HCN, PH$_3$ and H$_2$O quenching. They used the \citet{Zahnle2014} parameterization to find quench pressures ($P_{q,i}$ for species $i$) for all species (e.g., where $\tau_\mathrm{CO} = \tau_\mathrm{mix}$), then assumed chemical equilibrium for $P > P_{q,i}$, and constant volume mixing ratios for $P < P_{q,i}$ with a concentration predicted by equilibrium chemistry at $P_{q,i}$.

This approach works reasonably well for many species, but not CO$_2$. CO quenches at a higher pressure than CO$_2$, so even after CO has quenched, the CO$_2$ abundance is determined by the net equilibrium reaction: 
\begin{equation} \label{eq:co2eq}
  \mathrm{CO_2} + \mathrm{H_2} \rightleftarrows \mathrm{CO} + \mathrm{H_2O},
\end{equation}
until CO$_2$ itself quenches with respect to the \textit{disequilibrium} (quenched) abundance of CO.
This chemical behavior and the corresponding CO$\rightleftarrows$CO$_2$ reaction kinetics were briefly described by \cite{Visscher2010} for Jupiter’s atmosphere, where secondary quenching has a minimal effect on the CO$_2$ abundance. The Sonora Elf Owl grid thus did not account for this effect, which appears to be more pronounced in late T and Y dwarfs -- the grid quenched CO$_2$ relative to full atmospheric equilibrium, resulting in mid- to upper-atmosphere CO$_2$ concentrations too small by over one order of magnitude for objects with a $T_\mathrm{eff} \lesssim 600$ K. As demonstrated by \cite{Beiler2024}, CO$_2$ features in the JWST spectra of late T and Y dwarfs with the Elf Owl atmospheres could only be fit by accounting for secondary CO$_2$ quenching with respect to CO.

\begin{figure*}
  \centering
  \includegraphics[width=1\textwidth]{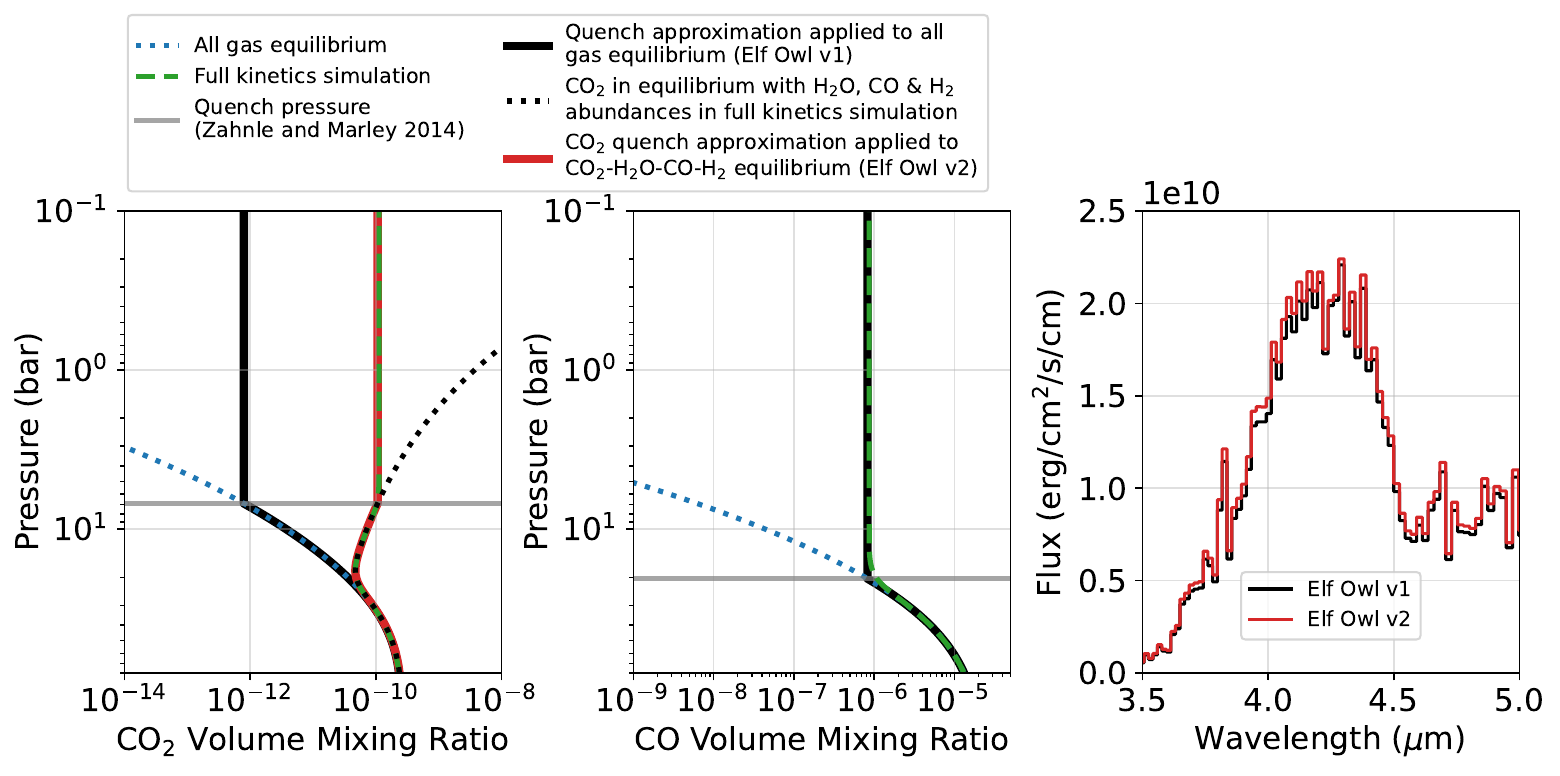}
  \caption{CO$_2$ (left) and CO (center) quenching in a brown dwarf atmosphere with $T_\mathrm{eff}$ = 500 K, a 31 m/s$^2$ gravity, $0.1\times$ solar metallicity, solar carbon-to-oxygen, and a vertically constant $K_{zz} = 10^8$ cm$^2$/s. The green dashed lines are the results of a full kinetics simulation using the \texttt{Photochem} code. CO (right) quenches with respect to the full atmosphere equilibrium (blue dotted line). CO$_2$ (left) quenches with respect to the already quenched abundances of CO (black dotted line). The Sonora Elf Owl v1 published in \citet{Mukherjee2024} quenched CO$_2$ with respect to the full atmosphere equilibrium, under predicting the gases abundance by two order of magnitude (black solid line, left panel). The right panel compares the emission spectra of Elf Owl v1 (with the CO$_2$ under-predicted) and Elf Owl v2 (with the CO$_2$ corrected) for this substellar atmosphere.}
  \label{fig:example}
\end{figure*}

Figure \ref{fig:example} further illustrates the problem in the Sonora Elf Owl for one of the simulated brown dwarfs in the grid with $T_\mathrm{eff}$ = 500 K, a 31 m/s$^2$ gravity, $0.1\times$ solar metallicity, solar carbon-to-oxygen, and a vertically constant $K_{zz} = 10^8$ cm$^2$/s. The dashed green lines are the results of a full-kinetics simulation using the \texttt{Photochem} code \citep[v0.6.2;][]{Wogan2023} that tracks $\sim600$ reversible reactions for $\sim100$ species composted of H, He, C, O, N, and S. For CO (Figure \ref{fig:example}, right), the \texttt{Photochem} simulation matches the full atmosphere chemical equilibrium at depth until 20 bars pressure, where the CO abundances quenches. The quench pressure predicted by the full kinetics model agrees well with the \citet{Zahnle2014} parameterization (grey line in the center panel). 

CO$_2$ quenching is more complicated than CO quenching. For CO$_2$ (Figure \ref{fig:example}, left), the \texttt{Photochem} result tracks the full atmosphere equilibrium for $P > 20$ bar, but then deviates at the CO quench pressure ($P = 20$ bar) because CO$_2$ remains in equilibrium with the quenched CO abundance. This is made clear by the black dotted line in Figure \ref{fig:example} (left), which is the CO$_2$ abundance in equilibrium with H$_2$O, CO and H$_2$ in the full kinetics simulation, computed with

\begin{equation} \label{eq:co2eq_1}
  f_\mathrm{CO_2} = K_\mathrm{eq} \frac{f_\mathrm{CO} f_\mathrm{H_2O}}{f_\mathrm{H_2}}
\end{equation}
Here, $f_i$ is the volume mixing ratio of species $i$, and $K_\mathrm{eq} = 18.3 \exp(-2376/T - (932/T)^2)$ is the equilibrium constant for Reaction \eqref{eq:co2eq}. In Figure \ref{fig:example} (left), the kinetics model predicts a CO$_2$ abundances that tracks the CO$_2$-H$_2$O-CO-H$_2$ equilibrium (Equation \eqref{eq:co2eq_1}) until the CO$_2$ quench pressure at 7 bar, freezing in a $10^{-10}$ CO$_2$ volume mixing ratio. The \citet{Mukherjee2024} Sonora Elf Owl grid instead quenched CO$_2$ with respect to the full atmosphere equilibrium predicting a $< 7$ bar CO$_2$ abundance too small by two orders of magnitude in this scenario (black solid line in Figure \ref{fig:example}, left).

A full kinetics model is not required to accurately predict the quenched CO$_2$ abundance in a brown dwarf atmosphere where CO$_2$ is minor compared to CO. To predict a CO$_2$ concentration, a model should first compute the quench pressure of CO/H$_2$O and CO$_2$ using the reaction timescales in \citet{Zahnle2014} (e.g., Equation \eqref{eq:coch4}). The full atmosphere equilibrium at the CO and H$_2$O quench pressure determines the quenched concentrations of both gases. Finally, the quenched CO$_2$ abundance is given by Equation \eqref{eq:co2eq_1} evaluated at the CO$_2$ quench pressure using the quenched abundances of CO and H$_2$O.

Using this approach, we have updated the CO$_2$ abundances for the \citet{Mukherjee2024} Sonora Elf Owl grid, and uploaded it to the following Zenodo archives as version 2: 
\begin{itemize}
    \item Y-type models: \dataset[DOI: 10.5281/zenodo.15150865]{https://doi.org/10.5281/zenodo.15150865}
    \item T-type models: \dataset[DOI: 10.5281/zenodo.15150874]{https://doi.org/10.5281/zenodo.15150874}
    \item L-type models: \dataset[DOI: 10.5281/zenodo.15150881]{https://doi.org/10.5281/zenodo.15150881}
\end{itemize}
The altered grid also contains updated emission spectra reflecting the new CO$_2$ abundances (e.g., Figure \ref{fig:example}, right). Another change is that the spectra do not include PH$_3$ opacity, because observations have revealed less PH$_3$ then our model previously predicted \citep{Beiler2024}. Sonora Elf Owl v2 no longer has a pressure-temperature profile that is strictly self-consistent with the composition, because we did not re-run the climate simulations to account for the new CO$_2$ abundances. However, given that CO$_2$ is very minor in many substellar atmospheres (e.g., 0.1 ppbv, Figure \ref{fig:example}), we expect the climate impact of the new CO$_2$ concentrations to be negligible in many cases. A full grid of self-consistent models incorporating the corrected CO$_2$ abundances will be released in a future publication.

\section*{Acknowledgements}

N.F.W was supported by the NASA Postdoctoral Program.

\bibliography{bib}

\begin{thebibliography}{}
\expandafter\ifx\csname natexlab\endcsname\relax\def\natexlab#1{#1}\fi
\providecommand{\url}[1]{\href{#1}{#1}}
\providecommand{\dodoi}[1]{doi:~\href{http://doi.org/#1}{\nolinkurl{#1}}}
\providecommand{\doeprint}[1]{\href{http://ascl.net/#1}{\nolinkurl{http://ascl.net/#1}}}
\providecommand{\doarXiv}[1]{\href{https://arxiv.org/abs/#1}{\nolinkurl{https://arxiv.org/abs/#1}}}

\bibitem[{{Beiler} {et~al.}(2024){Beiler}, {Mukherjee}, {Cushing}, {Kirkpatrick}, {Schneider}, {Kothari}, {Marley}, \& {Visscher}}]{Beiler2024}
{Beiler}, S.~A., {Mukherjee}, S., {Cushing}, M.~C., {et~al.} 2024, \apj, 973, 60, \dodoi{10.3847/1538-4357/ad6759}

\bibitem[{{Marley} {et~al.}(2021){Marley}, {Saumon}, {Visscher}, {Lupu}, {Freedman}, {Morley}, {Fortney}, {Seay}, {Smith}, {Teal}, \& {Wang}}]{Marley2021}
{Marley}, M.~S., {Saumon}, D., {Visscher}, C., {et~al.} 2021, \apj, 920, 85, \dodoi{10.3847/1538-4357/ac141d}

\bibitem[{{Mukherjee} {et~al.}(2024){Mukherjee}, {Fortney}, {Morley}, {Batalha}, {Marley}, {Karalidi}, {Visscher}, {Lupu}, {Freedman}, \& {Gharib-Nezhad}}]{Mukherjee2024}
{Mukherjee}, S., {Fortney}, J.~J., {Morley}, C.~V., {et~al.} 2024, \apj, 963, 73, \dodoi{10.3847/1538-4357/ad18c2}

\bibitem[{{Visscher} {et~al.}(2010){Visscher}, {Moses}, \& {Saslow}}]{Visscher2010}
{Visscher}, C., {Moses}, J.~I., \& {Saslow}, S.~A. 2010, \icarus, 209, 602, \dodoi{10.1016/j.icarus.2010.03.029}

\bibitem[{{Wogan} {et~al.}(2023){Wogan}, {Catling}, {Zahnle}, \& {Lupu}}]{Wogan2023}
{Wogan}, N.~F., {Catling}, D.~C., {Zahnle}, K.~J., \& {Lupu}, R. 2023, \psj, 4, 169, \dodoi{10.3847/PSJ/aced83}

\bibitem[{{Zahnle} \& {Marley}(2014)}]{Zahnle2014}
{Zahnle}, K.~J., \& {Marley}, M.~S. 2014, \apj, 797, 41, \dodoi{10.1088/0004-637X/797/1/41}

\end{thebibliography}
\bibliographystyle{aasjournal}

\end{document}